\pdfoutput=1
\documentclass[sigconf,nonacm,acmthm=false]{acmart}

\usepackage{booktabs}
\usepackage{placeins}
\usepackage{algorithm}
\usepackage{algpseudocode}
\algrenewcommand\algorithmicrequire{\textbf{Input:}}
\algrenewcommand\algorithmicensure{\textbf{Output:}}
\algtext*{EndFor}
\usepackage{amsmath,mathtools}   % acmart already provides the AMS symbols through newtxmath
\usepackage{amsthm}
\theoremstyle{plain}
\newtheorem{theorem}{Theorem}
\newtheorem{lemma}{Lemma}
\newtheorem{proposition}{Proposition}

\theoremstyle{definition}

\newtheorem{definition}{Definition}
\newtheorem{problem}{Problem}

\begin{document}

\title{Certified Zonotopic Reachability Analysis for Trajectory Tracking on Lie Groups}

\author{Yanliang Huang}
\affiliation{\institution{Technical University of Munich}\country{}}
\email{yanliang.huang@tum.de}
\author{Peng Xie}
\affiliation{\institution{Technical University of Munich}\country{}}
\email{p.xie@tum.de}
\author{Zhen Zhang}
\affiliation{\institution{Technical University of Munich}\country{}}
\email{zhenzhang.zhang@tum.de}
\author{Zhuoqi Zeng}
\affiliation{\institution{Hainan Bielefeld University of Applied Sciences}\country{}}
\email{zhuoqi.zeng@hibiuh.edu.cn}
\author{Amr Alanwar}
\affiliation{\institution{Technical University of Munich}\country{}}
\email{alanwar@tum.de}

\begin{abstract}
Autonomous vehicles track planned trajectories close to obstacles. Unknown disturbances make them deviate from these trajectories. With a bound on the tracking error that disturbances can cause, a planned trajectory can be certified as safe before execution. Existing set-based methods bound the tracking error by one invariant ellipsoid for the whole trajectory or by intervals, which are both conservative. For aerial and space vehicles, the tracking error evolves on a matrix Lie group with nonlinear dynamics. However, for group-affine systems, the error dynamics are linear in Lie algebra coordinates when the vehicle receives the inputs of the reference. The control input and the disturbance enter through the inverse right Jacobian, a matrix that depends on the error. We propagate the linear part of the error dynamics with zonotopes and enclose the effect of the inverse right Jacobian on the disturbance in closed form. The resulting tubes are tight and provably contain all reachable tracking errors in continuous time. When some error components have no input, the inverse right Jacobian leaves unmatched nonlinear terms in the closed loop. None of the tested methods certifies such a closed loop. Hybrid coordinates make these terms matched, so that our controller cancels every nonlinear term that does not contain the disturbance. Our tubes give tighter bounds than the ellipsoid method in simulations of a planar aircraft and a fully actuated multirotor and in real flights of a quadrotor, where the tube contains every measured attitude error. Under larger disturbances and the same feedback gain, the ellipsoid method yields no valid ellipsoid, whereas our tubes remain valid.
\end{abstract}


\maketitle

\section{Introduction}

% --- Section 1 Introduction, paragraph 1

Autonomous vehicles track planned trajectories close to obstacles with feedback controllers. Disturbances that are unknown before execution make the vehicle deviate from the planned trajectory. We call the planned trajectory the reference and the deviation from it the tracking error. A planned trajectory can be certified before execution with a tube, a time-varying set that contains every tracking error that disturbances within given bounds can cause at every time instant~\cite{chutinan2003}. If the region that the tube covers around the reference avoids all obstacles, the vehicle avoids them during execution. A tighter tube certifies trajectories that pass closer to obstacles.

% --- Section 1, paragraph 2

Reachability analysis propagates the set of reachable tracking errors through the error dynamics~\cite{althoff2008,cora}. The reachable sets at all times form the tube. When the attitude of an aerial vehicle is represented by Euler angles, trigonometric functions of these angles enter the error dynamics and make them nonlinear. Standard reachability methods can linearize these dynamics in every step and treat the linearization error as an additional uncertain input, whose bound grows with the reachable set~\cite{althoff2008}. Each enlargement of the reachable set then causes a larger enlargement in the next step.

% --- Section 1, paragraph 3

The tracking error $\bar X^{-1}X$ between the reference $\bar X$ and the state $X$ is an element of a matrix Lie group, such as the rotation group $SO(3)$ or the group $SE_2(3)$ of attitude, velocity and position. Euler angles with Cartesian positions are one choice of coordinates for the tracking error. Lie algebra coordinates are another choice, which represent the tracking error by its matrix logarithm written as a vector. Group-affine systems~\cite{barrau2017} are systems on matrix Lie groups whose dynamics satisfy an affine identity with respect to the group product, a class that includes rigid-body kinematics. For these systems, the error dynamics in Lie algebra coordinates are linear when the vehicle receives the inputs of the reference, whereas in Euler angles they remain nonlinear. The input deviations, including the disturbance, enter the error dynamics through the inverse of the right Jacobian, a matrix that depends on the error. The ellipsoid method of Lin et al.~\cite{lin2024loglinear} uses this structure and bounds the tracking error by a single invariant ellipsoid. It bounds all disturbance components jointly by one norm ball and the inverse right Jacobian by one bound that holds on the whole ellipsoid. The ellipsoid thus loses the componentwise bounds of the disturbance and keeps one size for the whole trajectory. The interval method of Harapanahalli and Coogan~\cite{harapanahalli2024} propagates boxes in the Lie algebra. These boxes are enlarged whenever the error dynamics rotate them. A tight tube requires a set representation that is exact under the linear part of the error dynamics and keeps the componentwise bounds of the disturbance. A tight tube also requires a bound on the effect of the inverse right Jacobian on the disturbance that is computed in every step and holds for every error in the current set. We propagate the tracking error in Lie algebra coordinates with zonotopes~\cite{girard2005}, which are closed under linear maps and Minkowski sums and therefore keep the componentwise bounds of the disturbance. In every step, we enclose the disturbance multiplied by the inverse right Jacobian in closed form, with a bound on the higher-order remainder of the series of the inverse right Jacobian that is attained in the worst case.

% （第 5 段：闭环与 hybrid 坐标，贡献 2）

In closed loop, the inverse right Jacobian also multiplies the control input, which the controller computes from the error, so the closed loop contains terms that are nonlinear in the error. A term is matched if it enters an error component on which an input acts, and the controller can cancel only matched terms. When some error components have no input, the inverse right Jacobian couples the input into these components. The resulting unmatched term is quadratic in the error, and its bound grows with the tube. None of the methods we evaluate certifies this closed loop. In hybrid coordinates, which keep the Lie algebra coordinates for the attitude error and express the translational errors in the body frame of the reference, every term that is nonlinear in the error is matched. The controller cancels all of these terms except the terms that contain the disturbance, which the tube encloses, so the closed loop is certified.

\begin{figure*}[t]
\centering
\includegraphics[width=\textwidth]{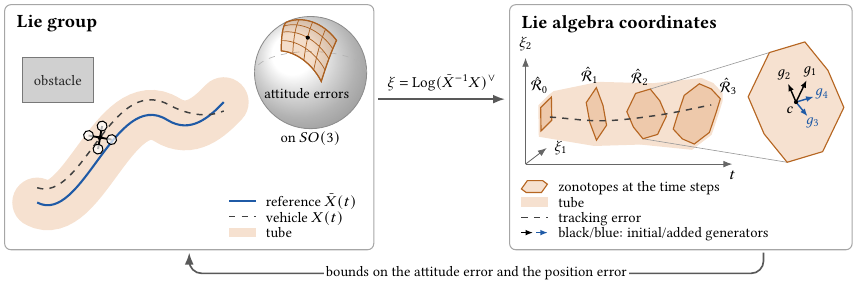}
\caption{The tracking error on the Lie group and in Lie algebra coordinates. In the left panel, the vehicle $X(t)$ follows the reference $\bar X(t)$, and the tube contains every tracking error that the disturbances can cause. A tube that avoids the obstacle certifies the reference. The inset shows a zonotope mapped onto $SO(3)$ by the exponential map, whose edges become curved, so we propagate the zonotopes in Lie algebra coordinates. The right panel shows zonotopes in Lie algebra or hybrid coordinates, which contain the tracking error at the time steps, and the tube, whose time-interval sets contain it between the time steps. The dashed curves are a trajectory of the vehicle and its tracking error. The enlarged zonotope consists of its center $c$ and its generators, and each step adds generators for the disturbance. The bounds on the attitude error and the position error are read from these sets.}
\label{fig:domains}
\end{figure*}

\subsection{Related Work}

% --- 1.1, paragraph 1

Tubes around a reference trajectory are a standard tool for motion planning under disturbances. Tube model predictive control bounds the deviation of a linear discrete-time system from its nominal trajectory by a disturbance invariant set~\cite{mayne2005robust,  zhang2026dtzpc}. Funnel libraries precompute funnels along a set of maneuvers by sum-of-squares programming and compose them online~\cite{majumdar2017funnel}. Control contraction metrics give tubes around every feasible nominal trajectory~\cite{singh2023contraction,zhao2022tube}. FaSTrack~\cite{chen2021fastrack} computes tracking error bounds that hold for every trajectory of a planner by Hamilton--Jacobi reachability. For quadrotors, geometric controllers on $SE(3)$ keep the tracking errors uniformly ultimately bounded under bounded disturbances~\cite{lee2013robust}. Zonotope reachability computes forward reachable sets of planned quadrotor trajectories with an approximation of the trajectory-dependent tracking error~\cite{kousik2019quadrotor}. Our tubes are reachable sets of the tracking error in Lie algebra coordinates, computed for one given reference. They contain the tracking error at every time instant.

% --- 1.1, paragraph 2

Set propagation computes over-approximations of reachable sets with zonotopes, support functions and other set representations~\cite{althoff2021set,zhang2025data}. Zonotopic reachability~\cite{girard2005} extends to nonlinear systems by conservative linearization~\cite{althoff2008}, which propagates the linearized dynamics with zonotopes and treats the linearization error as an additional uncertain input. Conservative polynomialization~\cite{althoff2013reachability} abstracts the dynamics to a polynomial differential inclusion and propagates polynomial zonotopes. Sparse polynomial zonotopes~\cite{kochdumper2021sparse} and the Taylor models of Flow*~\cite{chen2013flowstar} also represent nonconvex reachable sets. Adaptive choices of the time step and of the number of generators improve the accuracy~\cite{wetzlinger2021}. The toolboxes CORA~\cite{cora} and JuliaReach~\cite{bogomolov2019juliareach} implement these methods. Our method follows the step structure of conservative linearization. In the closed loops of this paper, the error dynamics in Lie algebra or hybrid coordinates are linear except for the terms that contain the disturbance. The linear part is propagated exactly, and a box encloses only these terms.

% --- 1.1, paragraph 3

Barrau and Bonnabel~\cite{barrau2017} introduced group-affine systems for state estimation. Lin et al.~\cite{lin2024loglinear} derived the error dynamics with control and disturbance inputs and designed the dynamic inversion controller. Woodbury et al.~\cite{condie2025} derived logarithmic error dynamics for thrusting spacecraft on $SE_2(3)$ with a bound on the gravity mismatch. Barfoot and Furgale~\cite{barfoot2014} and Barfoot~\cite{barfoot2024} give closed forms and the block structure of the right Jacobian and its inverse. Our remainder bound builds on this block structure. In later work, Lin et al.~\cite{lin2025backstepping} define a group error on $SE_2(3)$ that contains the velocity and position errors in the body frame of the reference, design a backstepping controller in its logarithmic coordinates, and prove exponential stability of the error dynamics on one branch of the matrix logarithm. We use these body-frame errors for a static feedback under which the closed-loop error dynamics have the form that our tube computation requires.

% --- 1.1, paragraph 4

The ellipsoid method has been extended to multirotors on $SE_2(3)$~\cite{lin2023multirotor}. The interval method recenters its boxes with a truncated Baker--Campbell--Hausdorff series~\cite{harapanahalli2024}. For uncertain initial states, contraction analysis yields metric balls~\cite{harapanahalli2025,xu2025} and interval tubes exploit Lie symmetries~\cite{damers2022}. Sampling-based methods construct geodesic balls of attitude sets~\cite{faraci2025}. For state estimation, a zonotopic set-membership filter propagates the error in Lie algebra coordinates with a first-order linearization~\cite{li2025inzsmf}. Boxes on Lie groups support static set inversion~\cite{merlinge2024}. We compute continuous-time tubes under persistent disturbances and bound all higher-order terms.

\subsection{Contributions}

We summarize our contributions as follows.

\begin{itemize}
\item A zonotope-based reachability algorithm for the tracking errors of group-affine systems on $SO(3)$, $SE(2)$ and the groups $SE_q(3)$ in Lie algebra coordinates. The algorithm accumulates the disturbance in the generators of the zonotope and encloses the effect of the inverse right Jacobian on the disturbance in closed form. The tube of the algorithm contains the tracking error at every time instant.
\item Hybrid coordinates in which a static feedback cancels the nonlinear terms without the disturbance, so that the algorithm certifies closed loops with unmatched nonlinear terms.
\item An evaluation on a planar aircraft, the attitude benchmark of the interval method, a fully actuated multirotor and flights of a real quadrotor. The tubes are tighter than the bounds of the ellipsoid method and the interval method. They remain valid under disturbances for which the ellipsoid method yields no valid ellipsoid for the same feedback gain and a standard reachability method in Cartesian coordinates with Euler angles loses its guarantee.
\end{itemize}

\section{Preliminaries and Problem Formulation}

\subsection{Notations}
For a vector $a\in\mathbb R^n$, $a^{(i)}$ denotes its $i$-th element, a subscript that names a part of $a$, such as its rotation part, denotes the block of elements of this part, and $A^\top$ denotes the transpose of a matrix $A$. Absolute values and inequalities of vectors and matrices are taken elementwise, $|A|^2$ denotes $|A||A|$, $\|\cdot\|$ is the Euclidean norm and $\mathbf 1$ is the vector of ones. The box with half-widths $v\ge0$ is $\mathrm{Box}(v)=\{x:|x|\le v\}$, the Minkowski sum of two sets is $\mathcal A\oplus\mathcal B=\{a+b:a\in\mathcal A,\,b\in\mathcal B\}$, and $\mathrm{conv}(\cdot)$ denotes the convex hull.

\subsection{Preliminaries}

\begin{definition}[Matrix Lie group]
A matrix Lie group $\mathcal G$ is a closed subgroup of the group of invertible real $m\times m$ matrices with matrix multiplication as group operation. Its Lie algebra $\mathfrak g=\{\Xi\in\mathbb R^{m\times m}\mid\exp(t\Xi)\in\mathcal G\ \text{for all}\ t\in\mathbb R\}$, with the matrix exponential $\exp$, is a vector space of dimension $n$.
\end{definition}

The groups in this paper are $SO(3)$, $SE(2)$ and the groups $SE_q(3)$ of the matrices $\begin{bmatrix}R&P\\0&I_q\end{bmatrix}$ with $R\in SO(3)$ and $q$ translational vectors in $P\in\mathbb R^{3\times q}$, which for $SE_2(3)$ are the velocity and the position. The map $(\cdot)^\wedge:\mathbb R^n\to\mathfrak g$ and its inverse $(\cdot)^\vee$ identify $\mathfrak g$ with $\mathbb R^n$. The exponential map $\mathrm{Exp}(\xi)=\exp(\xi^\wedge)$ maps $\mathbb R^n$ to $\mathcal G$, and $\mathrm{Log}$ denotes the inverse of $\mathrm{Exp}$ on $\{\mathrm{Exp}(\xi):\theta<\pi\}$, where $\theta$ is the norm of the rotation part of $\xi$. The adjoint matrix $\xi^\curlywedge\in\mathbb R^{n\times n}$ is defined by $\xi^\curlywedge y=[\xi^\wedge,y^\wedge]^\vee=(\xi^\wedge y^\wedge-y^\wedge\xi^\wedge)^\vee$ for $y\in\mathbb R^n$ and defines the right Jacobian $J_r(\xi)=\sum_{k\ge0}(-\xi^\curlywedge)^k/(k+1)!$~\cite{barfoot2024}, which gives the time derivative of the exponential map, $\frac{d}{dt}\mathrm{Exp}(\xi)=\mathrm{Exp}(\xi)\,\bigl(J_r(\xi)\dot\xi\bigr)^\wedge$.

\begin{definition}[Group-affine system~\cite{barrau2017}]\label{def:groupaffine}
A system $\dot X=f_t(X)$ on $\mathcal G$ is group-affine if $f_t(X_1X_2)=f_t(X_1)X_2+X_1f_t(X_2)-X_1f_t(I)X_2$ for all $X_1,X_2\in\mathcal G$ and all $t$.
\end{definition}

Rigid-body kinematics driven by body velocities on $SO(3)$, $SE(2)$ and $SE_q(3)$, and the inertial navigation equations on $SE_2(3)$, are group-affine~\cite{barrau2017}. Barrau and Bonnabel~\cite{barrau2017} showed that, for two solutions $\bar X$ and $X$ of a group-affine system, the Lie algebra coordinates $\xi=\mathrm{Log}(\bar X^{-1}X)^\vee$ of the error between the two solutions satisfy a linear equation $\dot\xi=A(t)\xi$, which is the log-linear property.

Zonotopes represent the sets of tracking errors.

\begin{definition}[Zonotope~\cite{kuhn1998rigorously}]\label{def:zonotope}
Given a center $c\in\mathbb R^d$ and $\gamma\in\mathbb N$ generator vectors in a generator matrix $G=[g^{(1)}\ \cdots\ g^{(\gamma)}]\in\mathbb R^{d\times\gamma}$, a zonotope is the set
\begin{equation}\label{eq:zonotope}
  \mathcal Z=\Bigl\{x\in\mathbb R^d\ \Big|\ x=c+\sum_{i=1}^{\gamma}\beta^{(i)}g^{(i)},\ -1\le\beta^{(i)}\le1\Bigr\}.
\end{equation}
We use the shorthand notation $\mathcal Z=\langle c, G\rangle$.
\end{definition}

For a matrix $L\in\mathbb R^{d'\times d}$, the linear map of a zonotope is $L\mathcal Z=\langle Lc, LG\rangle$, and the Minkowski sum of two zonotopes $\mathcal Z_i=\langle c_i,G_i\rangle$ is $\mathcal Z_1\oplus\mathcal Z_2=\langle c_1+c_2, [G_1\ G_2]\rangle$~\cite{girard2005}. A box is the zonotope $\mathrm{Box}(v)=\langle 0, \mathrm{diag}(v)\rangle$, so disturbances with bounded components fit this representation.

\subsection{Problem Formulation}

Consider a vehicle with state $X\in\mathcal G$ that tracks a reference $\bar X\in\mathcal G$ on $[0,T]$,
\begin{equation}\label{eq:system}
  \dot{\bar X}=f_t(\bar X),\qquad \dot X=f_t(X)+X\,\bigl(B(u+w)\bigr)^\wedge,
\end{equation}
where the group-affine vector field $f_t$ contains the reference inputs, $u\in\mathbb R^{n_u}$ is the control input, $w\in\mathbb R^{n_u}$ is the disturbance and the matrix $B\in\mathbb R^{n\times n_u}$ maps the two inputs to the components of the Lie algebra on which the inputs act. The tracking error is $\xi=\mathrm{Log}(\bar X^{-1}X)^\vee$, and $\theta$ is the attitude error angle. The feedback controller is a dynamic inversion controller, which cancels the nonlinearity of the tracking error dynamics on the control input~\cite{lin2024loglinear}, with a given gain matrix $K$.

The disturbance is any measurable signal with values in the box $\mathcal Z_w=\langle0,\mathrm{diag}(\sigma)\rangle$ with half-widths $\sigma\ge0$, and the initial tracking error lies in a zonotope $\mathcal X_0$. We divide the horizon $[0,T]$ into $N$ steps of length $h=T/N$ with the times $t_k=kh$. A tube is a sequence of sets $\hat{\mathcal R}_{[k]}$, $k=0,\dots,N-1$, such that $\hat{\mathcal R}_{[k]}$ contains the tracking error on $[t_k,t_{k+1}]$ for every initial tracking error in $\mathcal X_0$ and every disturbance~\cite{althoff2008}.

\begin{problem}\label{prob:cert}
Given the reference $\bar X$, the gain matrix $K$, the disturbance box $\mathcal Z_w$ and the set $\mathcal X_0$ of initial tracking errors, compute a tube and, from the tube, bounds over $[0,T]$ on the attitude error angle $\theta$ and, for groups with translational parts, on the distance between the positions of the vehicle and the reference.
\end{problem}

\section{Method}

% --- opening paragraph (roadmap)

We compute the tube for the tracking error. In Lie algebra coordinates, the tracking error dynamics are linear except for the terms in which the inverse right Jacobian multiplies the inputs, because this matrix depends on the tracking error. We propagate the linear part with a zonotope and accumulate the disturbance input in its generators. We bound the remaining terms in closed form on an a priori enclosure of the tracking error, a box that contains the tracking error during a time step. From the propagated zonotopes and these bounds, we compute the tube and prove that it contains the tracking error. When some error components have no input, the closed loop has unmatched nonlinear terms. In hybrid coordinates these terms become matched, so the controller cancels them and the same tube computation applies.

\subsection{Closed-Loop Error Dynamics}

For group-affine systems~\cite{barrau2017}, the Lie algebra coordinates $\xi$ of the tracking error satisfy
\begin{equation}\label{eq:loglin}
  \dot\xi=A(t)\,\xi+J_r^{-1}(\xi)\,B(u+w)
\end{equation}
while $\theta<\pi$, where $J_r(\xi)$ is invertible. The group-affine property gives
\[
  \frac{d}{dt}(\bar X^{-1}X)=f_t(\bar X^{-1}X)-f_t(I)\,\bar X^{-1}X+\bar X^{-1}X\bigl(B(u+w)\bigr)^\wedge,
\]
and the log-linear property~\cite{barrau2017} gives
\[
  f_t(\mathrm{Exp}(\xi))-f_t(I)\,\mathrm{Exp}(\xi)=\mathrm{Exp}(\xi)\bigl(J_r(\xi)A(t)\xi\bigr)^\wedge.
\]
The time derivative of $\mathrm{Exp}$ then yields~\eqref{eq:loglin}~\cite{lin2024loglinear}. The first term of~\eqref{eq:loglin} is linear in $\xi$, with a matrix $A(t)$ that depends on time through the reference inputs. The second term is nonlinear in $\xi$ because the inverse right Jacobian multiplies both the control input and the disturbance. The controller chooses the control input and can therefore cancel the inverse right Jacobian on it. When every error component has an input, $B$ is invertible, and the dynamic inversion controller $u=B^{-1}J_r(\xi)K\xi$ can achieve this cancellation~\cite{lin2024loglinear}. The closed loop
\begin{equation}\label{eq:inverted}
  \dot\xi=\bigl(A(t)+K\bigr)\xi+J_r^{-1}(\xi)Bw
\end{equation}
is therefore nonlinear only in the disturbance term.

The disturbance term splits by its order in the tracking error into three parts, which we treat differently. The inverse right Jacobian has the series~\cite{barfoot2024}
\begin{equation}\label{eq:jrinv}
  J_r^{-1}(\xi)=I+\tfrac12\xi^\curlywedge+\sum_{m\ge1}\frac{\mathrm B_{2m}}{(2m)!}(\xi^\curlywedge)^{2m}
\end{equation}
with the Bernoulli numbers $\mathrm B_{2m}$, and its first two terms give
\begin{equation}\label{eq:split}
  J_r^{-1}(\xi)Bw=Bw+\sum_jw^{(j)}\Lambda_j\xi+\tau(\xi,Bw),\qquad\Lambda_j=-\tfrac12(Be_j)^\curlywedge,
\end{equation}
where $w^{(j)}$ is the $j$-th component of the disturbance and $e_j$ the $j$-th unit vector. The first part, the input $Bw$, does not depend on the tracking error and accumulates in the generators of the zonotope. The second part, the bilinear term, is linear in the tracking error for each disturbance component and dominates when the error is small, and we bound it componentwise. The third part, the remainder, contains the terms of second and higher order in the tracking error, and we bound its norm in closed form. With this split, the closed loop becomes
\begin{equation}\label{eq:closedloop}
  \dot\xi=\underbrace{\bigl(A(t)+K\bigr)\xi}_{\text{linear}}+\underbrace{Bw}_{\text{input}}+\underbrace{\textstyle\sum_jw^{(j)}\Lambda_j\xi}_{\text{bilinear}}+\underbrace{\tau(\xi,Bw)}_{\text{remainder}},
\end{equation}
with $w(t)\in\mathcal Z_w$, which we call the closed-loop error equation.

\subsection{Bounds on the Nonlinear Terms}

On each time step, the bilinear term and the remainder of the closed-loop error equation~\eqref{eq:closedloop} depend on the tracking error. We bound both terms on a box $\mathrm{Box}(R)$ that contains the tracking error during the step, an a priori enclosure~\cite{nedialkov1999}, and then compute the half-widths $R$ of this box.

For the bilinear term, we take absolute values of its factors. Since $|\xi|\le R$ and $|w^{(j)}|\le\sigma^{(j)}$ for every component of the disturbance, the bilinear term satisfies
\begin{equation}\label{eq:bilinear}
  \Bigl|\sum_j w^{(j)}\Lambda_j\xi\Bigr|\le\sum_j\sigma^{(j)}|\Lambda_j|\,R=\bar\Lambda R
\end{equation}
elementwise, with $\bar\Lambda=\sum_j\sigma^{(j)}|\Lambda_j|$.

For the remainder, we derive a bound from the series of the inverse right Jacobian that can be evaluated on the box. On $SO(3)$, the norm of the remainder is at most $\kappa(\theta)$ times the norm of $Bw$, with
\begin{equation}\label{eq:kappa}
  \kappa(\theta)=1-\tfrac{\theta}{2}\cot\tfrac{\theta}{2}.
\end{equation}
The function $\kappa$ is the sum of the absolute values of the series terms beyond the first two, and $\kappa(\theta)\approx\theta^2/12$ for small $\theta$.

\begin{lemma}[Remainder bound]\label{lem:remainder}
Let $\theta<\pi$, let $\eta\in\mathbb R^n$, and let $\tau=\tau(\xi,\eta)$.
\begin{enumerate}
\item On $SO(3)$, $\|\tau\|\le\kappa(\theta)\|\eta\|$.
\item On $SE_q(3)$ with $\xi=(\phi,\nu_1,\dots,\nu_q)$ and $\eta=(\eta_\phi,\eta_1,\dots,\eta_q)$, the rotation part satisfies $\|\tau_\phi\|\le\kappa(\theta)\|\eta_\phi\|$, and the translational parts satisfy $\|\tau_i\|\le\kappa(\theta)\|\eta_i\|+\kappa'(\theta)\|\nu_i\|\|\eta_\phi\|$.
\item On $SE(2)$ with $\xi=(\rho,\theta_\xi)$, $\theta=|\theta_\xi|$ and $\eta=(\eta_\rho,\eta_\theta)$, the rotation part of $\tau$ vanishes, and the translational part satisfies $\|\tau_\rho\|\le\kappa(\theta)\|\eta_\rho\|+\frac{\kappa(\theta)}{\theta}\|\rho\|\,|\eta_\theta|$, with $\kappa(\theta)/\theta:=0$ at $\theta=0$.
\end{enumerate}
For any given $\theta$ and norms of the parts, there are $\xi$ and $\eta$ for which all bounds hold with equality.
\end{lemma}

\begin{proof}

The remainder is a power series in $\xi^\curlywedge$, so we bound the parts of each power of $\xi^\curlywedge$ applied to $\eta$ by powers of $\theta$ and sum these bounds with the weights of the series. The series \eqref{eq:jrinv} converges for $\theta<2\pi$, because the eigenvalues of $\xi^\curlywedge$ on $SO(3)$, $SE(2)$ and $SE_q(3)$ are $0$ and $\pm i\theta$. Subtracting the first two terms gives
\begin{equation}\label{eq:tauseries}
  \tau(\xi,\eta)=\sum_{m\ge1}\frac{\mathrm B_{2m}}{(2m)!}(\xi^\curlywedge)^{2m}\eta.
\end{equation}
Since $(-1)^m\mathrm B_{2m}=-|\mathrm B_{2m}|$ for $m\ge1$, the function $\kappa$ and its derivative are the series
\begin{equation}\label{eq:kappaseries}
  \kappa(\theta)=\sum_{m\ge1}\frac{|\mathrm B_{2m}|}{(2m)!}\theta^{2m},\qquad \kappa'(\theta)=\sum_{m\ge1}\frac{2m\,|\mathrm B_{2m}|}{(2m)!}\theta^{2m-1}.
\end{equation}

\emph{Item 1, $SO(3)$.} Here $\xi^\curlywedge=\phi^\wedge$, and $\|\phi^\wedge y\|\le\theta\|y\|$ for every $y\in\mathbb R^3$, so
\begin{equation}\label{eq:so3norm}
  \|(\xi^\curlywedge)^{2m}\eta\|\le\theta^{2m}\|\eta\|,
\end{equation}
and the weighted sum gives $\|\tau\|\le\kappa(\theta)\|\eta\|$. On the plane orthogonal to $\phi$, $(\phi^\wedge)^2=-\theta^2I$, so for $\eta\perp\phi$ the series sums to $\tau=-\kappa(\theta)\eta$, which attains the bound.

\emph{Item 2, $SE_q(3)$.} The adjoint acts on $(a,b_1,\dots,b_q)$ as
\begin{equation}\label{eq:adse}
  \xi^\curlywedge(a,b_1,\dots,b_q)=(\phi^\wedge a,\ \nu_1^\wedge a+\phi^\wedge b_1,\ \dots,\ \nu_q^\wedge a+\phi^\wedge b_q).
\end{equation}
By induction over $p$, the rotation part of $(\xi^\curlywedge)^p\eta$ is $(\phi^\wedge)^p\eta_\phi$, and part $i$ is
\begin{equation}\label{eq:adpower}
  \bigl((\xi^\curlywedge)^p\eta\bigr)_i=(\phi^\wedge)^p\eta_i+\sum_{l=0}^{p-1}(\phi^\wedge)^l\,\nu_i^\wedge\,(\phi^\wedge)^{p-1-l}\eta_\phi .
\end{equation}
With $\|\phi^\wedge y\|\le\theta\|y\|$, its norm is bounded by
\begin{equation}\label{eq:adpowernorm}
  \bigl\|\bigl((\xi^\curlywedge)^p\eta\bigr)_i\bigr\|\le\theta^p\|\eta_i\|+p\,\theta^{p-1}\|\nu_i\|\|\eta_\phi\|,
\end{equation}
and the weighted sum over $p=2m$ gives both bounds of item 2. Taking $\nu_i=\alpha_i\phi$, $\eta_\phi\perp\phi$ and $\eta_i=\beta_i\operatorname{sgn}(\alpha_i)\eta_\phi$ with $\beta_i\ge0$ makes $\nu_i^\wedge$ commute with $\phi^\wedge$, so both terms of $\tau_i$ are nonnegative multiples of $-\operatorname{sgn}(\alpha_i)\eta_\phi$, and all bounds hold with equality for any given norms.

\emph{Item 3, $SE(2)$.} With the rotation $J$ by $\pi/2$, the adjoint acts on $(y,z)\in\mathbb R^2\times\mathbb R$ as $\xi^\curlywedge(y,z)=(\theta_\xi Jy-zJ\rho,\,0)$, and by induction, for $p\ge1$,
\begin{equation}\label{eq:adse2}
  (\xi^\curlywedge)^p\eta=\bigl((\theta_\xi J)^p\eta_\rho-(\theta_\xi J)^{p-1}J\rho\,\eta_\theta,\ 0\bigr).
\end{equation}
With $(\theta_\xi J)^{2m}=(-\theta^2)^mI$ and $(\theta_\xi J)^{2m-1}J=(-1)^m\theta_\xi^{2m-1}I$, the series sums to
\begin{equation}\label{eq:tause2}
  \tau_\rho=-\kappa(\theta)\eta_\rho+\frac{\kappa(\theta)}{\theta_\xi}\eta_\theta\,\rho .
\end{equation}
The triangle inequality gives item 3, with equality when $\eta_\rho$ is a nonnegative multiple of $-\operatorname{sgn}(\theta_\xi\eta_\theta)\rho$, because both terms then point in the same direction.
\end{proof}

Lemma~\ref{lem:remainder} bounds the remainder by the attitude error angle and the norms of the parts of the tracking error and the disturbance $\eta=Bw$. The bound is attained, so it is the smallest bound in terms of these quantities. To use it on a step, we insert the largest values of these quantities on the box, which are the norms of the corresponding parts of the half-widths $R$ and of the disturbance bound $|B|\sigma$. The functions in Lemma~\ref{lem:remainder} are nondecreasing, so this bounds each part of the remainder for every tracking error in the box and every disturbance. Each element of a part is bounded by the norm of the part, and we denote the resulting elementwise bound by $\bar\tau(R)$.

The rate that~\eqref{eq:closedloop} allows on the box decides whether the box is an a priori enclosure.

\begin{lemma}[A priori enclosure]\label{lem:apriori}
Let $|\xi(t_k)|\le r_k$, let $A_k=A(t_k+h/2)+K$, let $D_k$ bound $|A(t)-A(t_k+h/2)|$ on the step, and let $F_k(R)=(|A_k|+D_k)R+|B|\sigma+\bar\Lambda R+\bar\tau(R)$. If $r_k+hF_k(R)\le R$ and every point of $\mathrm{Box}(R)$ satisfies $\theta<\pi$, then $|\xi(t)|\le R$ for all $t\in[t_k,t_{k+1}]$ and every disturbance with values in $\mathcal Z_w$.
\end{lemma}

\begin{proof}

The condition $\theta<\pi$ keeps the box in the domain of $\mathrm{Log}$, on which~\eqref{eq:closedloop} and Lemma~\ref{lem:remainder} hold. By the Carathéodory theorem,~\eqref{eq:closedloop} has a solution, which is unique because the right-hand side of~\eqref{eq:closedloop} is locally Lipschitz in $\xi$. For every continuous $y$ on the step with values in $\mathrm{Box}(R)$, the right-hand side of~\eqref{eq:closedloop} along $y$ is bounded elementwise by $F_k(R)$. Since $r_k+hF_k(R)\le R$, the Picard operator, which maps $y$ to $\xi(t_k)$ plus the integral of this right-hand side, returns a function with values in $\mathrm{Box}(R)$. The returned functions are equicontinuous, so the operator has a fixed point by Schauder's theorem, and by uniqueness this fixed point is the tracking error.
\end{proof}

We find such a box by the Picard iteration $R\leftarrow r_k+hF_k(R)$, with a slight enlargement of each iterate. We denote the box found on the step by $\mathrm{Box}(R_k)$.

\subsection{Tube Computation}

\begin{figure}
\centering
\includegraphics[width=\columnwidth]{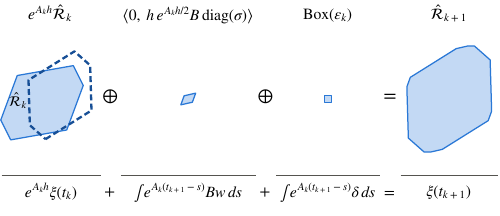}
\caption{One step of Algorithm~\ref{alg:tube}, with illustrative sets. The lower row is the solution~\eqref{eq:sol} of the closed-loop error equation over the step. The linear part maps the zonotope to another zonotope without over-approximation, the input adds a zonotope with one generator per disturbance component, and a box encloses the remaining terms. The sum of the three sets contains the tracking error at the end of the step.}\label{fig:terms}
\end{figure}

Let $\hat{\mathcal R}_k$ be a zonotope that contains the tracking error at $t_k$. On the step $[t_k,t_{k+1}]$, we collect the deviation of $A(t)$ from its midpoint value, the bilinear term and the remainder in $\delta$, so that the closed-loop error equation~\eqref{eq:closedloop} reads $\dot\xi=A_k\xi+Bw+\delta$ and its solution is
\begin{equation}\label{eq:sol}
  \begin{aligned}\xi(t_{k+1})={}&e^{A_kh}\xi(t_k)+\int_{t_k}^{t_{k+1}}e^{A_k(t_{k+1}-s)}Bw\,ds\\&+\int_{t_k}^{t_{k+1}}e^{A_k(t_{k+1}-s)}\delta\,ds.\end{aligned}
\end{equation}
Each term lies in a set, as Figure~\ref{fig:terms} shows. The first term lies in $e^{A_kh}\hat{\mathcal R}_k$, which is again a zonotope, so the linear part adds no over-approximation. A box, in contrast, would have to be enclosed in a larger box after a rotation, which is the wrapping effect~\cite{kuhn1998rigorously}. In the second term, replacing the propagator $e^{A_k(t_{k+1}-s)}$ by its midpoint value $e^{A_kh/2}$ turns the integral into $h\,e^{A_kh/2}B\bar w$, where the mean disturbance $\bar w$ over the step satisfies $|\bar w|\le\sigma$. This term therefore lies in the zonotope $\langle0,h\,e^{A_kh/2}B\,\mathrm{diag}(\sigma)\rangle$. The third term satisfies $|\delta|\le\bar\delta_k=D_kR_k+\bar\Lambda R_k+\bar\tau(R_k)$ on the a priori enclosure. Together with the error of replacing the propagator, it lies in a box, which we denote by $\mathrm{Box}(\varepsilon_k)$. Hence $\xi(t_{k+1})$ lies in the Minkowski sum of the three sets, which is again a zonotope.

Between $t_k$ and $t_{k+1}$, the rate of change of the tracking error is bounded on the a priori enclosure. The tracking error therefore stays close to the segment between its values at both ends, and a box bounds this deviation. The convex hull of $\hat{\mathcal R}_k$ and $\hat{\mathcal R}_{k+1}$ contains the segment, and enlarging it by this box gives the time-interval set $\hat{\mathcal R}_{[k]}$, which contains the tracking error on the whole step~\cite{althoff2008}.

\begin{lemma}[One step]\label{lem:step}
Let $\xi(t_k)\in\hat{\mathcal R}_k$ with $|\xi(t_k)|\le r_k$, let $\mathrm{Box}(R_k)$ satisfy the conditions of Lemma~\ref{lem:apriori}, and let
\begin{align*}
  &E=e^{|A_k|h},\quad \bar f_k=|B|\sigma+\bar\delta_k,\quad \varepsilon_k=hE\bar\delta_k+\tfrac{h^2}{4}|A_k|E|B|\sigma,\\
  &\varepsilon_{[k]}=\tfrac{h^2}{8}|A_k|^2E\,r_k+\bigl(\tfrac h2+\tfrac{h^2}{4}|A_k|\bigr)E\,\bar f_k,
\end{align*}
with $\bar\delta_k=D_kR_k+\bar\Lambda R_k+\bar\tau(R_k)$. Then, for every disturbance with values in $\mathcal Z_w$,
\begin{equation}\label{eq:steppoint}
  \begin{aligned}\xi(t_{k+1})\in{}&e^{A_kh}\hat{\mathcal R}_k\oplus\langle0,\,h\,e^{A_kh/2}B\,\mathrm{diag}(\sigma)\rangle\\&\oplus\mathrm{Box}(\varepsilon_k),\end{aligned}
\end{equation}
and, for every $t\in[t_k,t_{k+1}]$ and every set $\hat{\mathcal R}_{k+1}$ that contains $\xi(t_{k+1})$,
\begin{equation}\label{eq:stepinterval}
  \xi(t)\in\mathrm{conv}(\hat{\mathcal R}_k\cup\hat{\mathcal R}_{k+1})\oplus\mathrm{Box}(\varepsilon_{[k]}).
\end{equation}
\end{lemma}

\begin{proof}

We compare the tracking error with the solution of the linear system with the frozen matrix $A_k$ and bound the difference on the a priori enclosure. Shift time so that $t_k=0$. By Lemma~\ref{lem:apriori}, $|\xi|\le R_k$ on the step, so the term $\delta$ in~\eqref{eq:sol} satisfies $|\delta|\le\bar\delta_k$ and $|\dot\xi-A_k\xi|\le\bar f_k$. Three facts about the matrix exponential hold entrywise for $0\le s,a,b,t\le h$ and $\lambda=t/h$:
\begin{equation}\label{eq:expfacts}
  \begin{aligned}&|e^{A_ks}|\le E,\qquad |e^{A_ka}-e^{A_kb}|\le|a-b|\,|A_k|E,\\&\bigl|e^{A_kt}-(1-\lambda)I-\lambda e^{A_kh}\bigr|\le\frac{t(h-t)}{2}|A_k|^2E .\end{aligned}
\end{equation}
The first holds because $|A_k|$ is nonnegative, the second follows from $e^{A_ka}-e^{A_kb}=\int_b^aA_ke^{A_ks}ds$, and the third holds because every entry of $e^{A_kt}$ is a twice differentiable function of $t$ whose second derivative is bounded by the corresponding entry of $|A_k|^2E$.

For the time point, the input integral in~\eqref{eq:sol} splits at the midpoint propagator as
\begin{equation}\label{eq:inputsplit}
  \begin{aligned}&\int_0^he^{A_k(h-s)}Bw(s)\,ds=h\,e^{A_kh/2}B\bar w\\&\quad+\int_0^h\bigl(e^{A_k(h-s)}-e^{A_kh/2}\bigr)Bw(s)\,ds,\\&\bar w=\frac1h\int_0^hw(s)\,ds\in\mathcal Z_w .\end{aligned}
\end{equation}
The first term lies in $\langle0,\,h\,e^{A_kh/2}B\,\mathrm{diag}(\sigma)\rangle$. The second fact bounds the last integral by $|A_k|E|B|\sigma\int_0^h|h/2-s|\,ds=\frac{h^2}{4}|A_k|E|B|\sigma$, and the first fact bounds $\int_0^he^{A_k(h-s)}\delta(s)\,ds$ by $hE\bar\delta_k$. The sum of the two bounds is $\varepsilon_k$.

For the time interval, let $f=\dot\xi-A_k\xi$, $P=\int_0^te^{A_k(t-s)}f\,ds$ and $Q=\int_t^he^{A_k(h-s)}f\,ds$. Since $(1-\lambda)\xi(0)+\lambda\xi(h)$ lies in $\mathrm{conv}(\hat{\mathcal R}_k\cup\hat{\mathcal R}_{k+1})$, it suffices to bound
\begin{equation}\label{eq:segment}
  \begin{aligned}&\xi(t)-(1-\lambda)\xi(0)-\lambda\xi(h)\\&\ =\bigl(e^{A_kt}-(1-\lambda)I-\lambda e^{A_kh}\bigr)\xi(0)+\bigl(I-\lambda e^{A_k(h-t)}\bigr)P-\lambda Q .\end{aligned}
\end{equation}
The third fact bounds the first term by $\frac{h^2}{8}|A_k|^2Er_k$. With $|P|\le te^{|A_k|t}\bar f_k$, $|Q|\le(h-t)e^{|A_k|(h-t)}\bar f_k$ and $|e^{A_k(h-t)}-I|\le(h-t)|A_k|e^{|A_k|(h-t)}$, the remaining terms are bounded by $\bigl(\frac{2t(h-t)}{h}+\frac{t^2(h-t)}{h}|A_k|\bigr)E\bar f_k$, and $2t(h-t)/h\le h/2$ and $t^2(h-t)/h\le h^2/4$ on $[0,h]$ give $\varepsilon_{[k]}$.
\end{proof}

The time-interval sets of all steps form the tube. Algorithm~\ref{alg:tube} computes these sets step by step.

\begin{algorithm}
\caption{Tube computation}\label{alg:tube}
\begin{algorithmic}[1]
\Require set $\mathcal X_0$ of initial tracking errors, disturbance half-widths $\sigma$, system matrix $A(t)$, gain matrix $K$, input matrix $B$, bilinear matrices $\Lambda_j$, remainder bound $\bar\tau$, step $h$, horizon $T$, inflation $\mu>0$, iteration limit $i_{\max}$.
\Ensure zonotopes $\hat{\mathcal R}_k$ at the time steps and time-interval sets $\hat{\mathcal R}_{[k]}$ for $k=0,\dots,N_{\mathrm c}-1$ with $t_{N_{\mathrm c}}\le T$.
\State\label{ln:1} $\langle c_0,G_0\rangle\leftarrow\mathcal X_0$, $\hat{\mathcal R}_0\leftarrow\mathcal X_0$, $t_0\leftarrow0$, $k\leftarrow0$
\While{$t_k<T$}\label{ln:2}
\State\label{ln:3} $r_k\leftarrow|c_k|+|G_k|\mathbf 1$
\State\label{ln:4} $A_k\leftarrow A(t_k+h/2)+K$
\State\label{ln:5} choose $D_k$ with $|A(t)-A(t_k+h/2)|\le D_k$ for all $t\in[t_k,t_k+h]$
\State\label{ln:6} $R\leftarrow r_k+hF_k(r_k)$
\For{$i=1,\dots,i_{\max}$}\label{ln:7}
\State\label{ln:8} $\tilde R\leftarrow(1+\mu)R$
\State\label{ln:9} \textbf{if} $r_k+hF_k(\tilde R)\le\tilde R$ \textbf{then} $R_k\leftarrow\tilde R$ and exit the loop
\State\label{ln:10} $R\leftarrow\max\{R,\ r_k+hF_k(\tilde R)\}$
\EndFor
\State\label{ln:11} \textbf{if} no $R_k$ is found \textbf{or} $\mathrm{Box}(R_k)$ contains a point with $\theta\ge\pi$ \textbf{then return} with $N_{\mathrm c}=k$
\State\label{ln:12} $E\leftarrow e^{|A_k|h}$
\State\label{ln:13} $\bar\delta_k\leftarrow D_kR_k+\bar\Lambda R_k+\bar\tau(R_k)$, $\ \bar f_k\leftarrow|B|\sigma+\bar\delta_k$
\State\label{ln:14} $\varepsilon_k\leftarrow hE\bar\delta_k+\frac{h^2}{4}|A_k|E|B|\sigma$
\State\label{ln:15} $\varepsilon_{[k]}\leftarrow\frac{h^2}{8}|A_k|^2E\,r_k+\bigl(\frac h2+\frac{h^2}{4}|A_k|\bigr)E\,\bar f_k$
\State\label{ln:16} $c_{k+1}\leftarrow e^{A_kh}c_k$
\State\label{ln:17} $G_{k+1}\leftarrow\bigl[\,e^{A_kh}G_k\ \ \ h\,e^{A_kh/2}B\,\mathrm{diag}(\sigma)\ \ \ \mathrm{diag}(\varepsilon_k)\,\bigr]$
\State\label{ln:18} optionally $\langle c_{k+1},G_{k+1}\rangle\leftarrow\mathrm{reduce}(\langle c_{k+1},G_{k+1}\rangle)$
\State\label{ln:19} $\hat{\mathcal R}_{k+1}\leftarrow\langle c_{k+1},G_{k+1}\rangle$
\State\label{ln:20} $\hat{\mathcal R}_{[k]}\leftarrow\mathrm{conv}(\hat{\mathcal R}_k\cup\hat{\mathcal R}_{k+1})\oplus\mathrm{Box}(\varepsilon_{[k]})$
\State\label{ln:21} $t_{k+1}\leftarrow t_k+h$, $k\leftarrow k+1$
\EndWhile\label{ln:22}
\State\label{ln:23} $N_{\mathrm c}\leftarrow k$
\end{algorithmic}
\end{algorithm}

Lines~\ref{ln:6} to~\ref{ln:11} compute the a priori enclosure of Lemma~\ref{lem:apriori}, with the maximum in line~\ref{ln:10} taken elementwise. Lines~\ref{ln:12} to~\ref{ln:15} evaluate the bounds of Lemma~\ref{lem:step} on it. The function $\mathrm{reduce}$ in line~\ref{ln:18} returns a zonotope with fewer generators that contains its argument. It keeps the number of generators bounded over long horizons~\cite{kopetzki2017reduction,xie2026orthogonal}.

\begin{theorem}[Soundness]\label{thm:soundness}
Let Algorithm~\ref{alg:tube} return the sets up to the horizon $t_{N_{\mathrm c}}$. Then, for every initial tracking error in $\mathcal X_0$ and every disturbance with values in $\mathcal Z_w$, the tracking error exists on $[0,t_{N_{\mathrm c}}]$, satisfies $\theta<\pi$ there, and satisfies $\xi(t_{k+1})\in\hat{\mathcal R}_{k+1}$ and $\xi(t)\in\hat{\mathcal R}_{[k]}$ for all $t\in[t_k,t_{k+1}]$ and $k=0,\dots,N_{\mathrm c}-1$.
\end{theorem}

\begin{proof}

We apply Lemma~\ref{lem:apriori} and Lemma~\ref{lem:step} step by step. For $k=0$, $\xi(0)\in\mathcal X_0=\hat{\mathcal R}_0$. If $\xi(t_k)\in\hat{\mathcal R}_k$ and step $k$ passed line~\ref{ln:11} of Algorithm~\ref{alg:tube}, then $|\xi(t_k)|\le r_k$ by line~\ref{ln:3}, and $\mathrm{Box}(R_k)$ satisfies the conditions of Lemma~\ref{lem:apriori}. The tracking error therefore exists on the step with $\theta<\pi$. Lemma~\ref{lem:step} places $\xi(t_{k+1})$ in the set $\langle c_{k+1},G_{k+1}\rangle$ of line~\ref{ln:17}, and the reduction in line~\ref{ln:18} returns a superset, so $\xi(t_{k+1})\in\hat{\mathcal R}_{k+1}$. Lemma~\ref{lem:step} also places $\xi(t)$ in $\hat{\mathcal R}_{[k]}$ for $t\in[t_k,t_{k+1}]$.
\end{proof}

Theorem~\ref{thm:soundness} solves Problem~\ref{prob:cert} up to $t_{N_{\mathrm c}}$ once the two norms are read from the tube. The attitude error angle is the norm of the rotation part of $\xi$. The distance between the positions of the vehicle and the reference is at most the norm of the position part, because the position difference is this part multiplied by the reference rotation and by a matrix of operator norm at most one. For each part, we bound its norm over a zonotope by covering the unit sphere with caps, dividing the support function~\cite{leguernic2010} at each cap center by the cosine of the cap radius and taking the largest quotient. Over $[t_k,t_{k+1}]$, we take the larger of the bounds on $\hat{\mathcal R}_k$ and $\hat{\mathcal R}_{k+1}$ plus the norm of that part of $\varepsilon_{[k]}$, because a norm is convex.

\subsection{Hybrid Coordinates}

The controller of Section 3.1 requires an input on every error component, so that every nonlinear term is matched. For vehicles on $SE_2(3)$ whose inputs are the angular rate and the specific force, such as a fully actuated multirotor, the position is driven only through the velocity, so $B$ has no inverse. We change the error coordinates so that a static feedback cancels every nonlinear term that does not contain the disturbance.

Such a vehicle has the attitude $R$, the velocity $v$ and the position $p$, which satisfy
\begin{equation}\label{eq:model}
  \dot R=R\,(\bar\omega+u_\omega+w_g)^\wedge,\qquad \dot v=R(\bar a+u_a+w_a)+\mathsf g,\qquad\dot p=v,
\end{equation}
with the body angular rate $u_\omega$ and the specific force $u_a$ as inputs, the disturbances $w_g$ and $w_a$ on these inputs, bounded componentwise by $\sigma_g$ and $\sigma_a$, and gravity $\mathsf g$. The reference $(\bar R,\bar v,\bar p)$ follows the same equations with $u=0$, $w=0$ and given smooth inputs $\bar\omega(t)$ and $\bar a(t)$. Let $\bar A(\bar\omega,\bar a)$ denote the matrix with the blocks $-\bar\omega^\wedge$ on the diagonal, $-\bar a^\wedge$ in the velocity row and attitude column and $I$ in the position row and velocity column, let $B_c=[I_6\ 0]^\top$, and let $K\in\mathbb R^{6\times9}$ be the gain.

For these vehicles, the Lie algebra coordinates of the velocity and position errors depend on the attitude error, so the inverse right Jacobian couples the angular rate input into the position row. The feedback that makes the attitude and velocity rows linear then leaves an unmatched term in the position row that is quadratic in the error. We therefore keep the Lie algebra coordinate $\phi$ only for the attitude and express the velocity and position errors in the body frame of the reference. This frame does not depend on the tracking error, so every term that is nonlinear in the error is matched. We call
\begin{equation}\label{eq:hybrid}
  x=(\phi,\nu,\rho),\quad \phi=\mathrm{Log}(\bar R^\top R)^{\vee},\quad \nu=\bar R^\top(v-\bar v),\quad \rho=\bar R^\top(p-\bar p),
\end{equation}
the hybrid coordinates. The attitude part of $x$ equals the attitude part of $\xi$, so $\theta=\|\phi\|$, and $\|\rho\|=\|p-\bar p\|$ because $\bar R$ preserves the norm.

\begin{proposition}[Closed loop in hybrid coordinates]\label{prop:hybrid}
While $\theta<\pi$, the feedback
\begin{equation}\label{eq:feedback}
  u_\omega=J_r(\phi)\,(Kx)_\omega,\quad u_a=\mathrm{Exp}(\phi)^\top\bigl[(Kx)_a-(\mathrm{Exp}(\phi)-I)\,\bar a-\bar a^\wedge\,\phi\bigr]
\end{equation}
gives
\begin{equation}\label{eq:hybridloop}
  \dot x=\bigl(\bar A(\bar\omega(t),\bar a(t))+B_cK\bigr)x+\begin{bmatrix}J_r^{-1}(\phi)\,w_g\\ \mathrm{Exp}(\phi)\,w_a\\ 0\end{bmatrix}.
\end{equation}
\end{proposition}

\begin{proof}

We differentiate the three parts of $x$ and insert the feedback. For the attitude, $\bar R^\top R=\mathrm{Exp}(\phi)$ and the attitude equations in~\eqref{eq:model} give
\begin{equation}\label{eq:attitude}
  J_r(\phi)\dot\phi=\bigl(I-\mathrm{Exp}(-\phi)\bigr)\bar\omega+u_\omega+w_g .
\end{equation}
The series of $J_r$~\cite{barfoot2024} gives the identity $J_r(\phi)\phi^\wedge=I-\mathrm{Exp}(-\phi)$, hence $J_r^{-1}(\phi)(I-\mathrm{Exp}(-\phi))\bar\omega=-\bar\omega^\wedge\phi$, and inserting $u_\omega$ gives $\dot\phi=-\bar\omega^\wedge\phi+(Kx)_\omega+J_r^{-1}(\phi)w_g$. For the velocity, gravity cancels and
\begin{equation}\label{eq:velocity}
  \dot\nu=-\bar\omega^\wedge\nu+\bigl(\mathrm{Exp}(\phi)-I\bigr)\bar a+\mathrm{Exp}(\phi)u_a+\mathrm{Exp}(\phi)w_a ,
\end{equation}
and inserting $u_a$ gives $\dot\nu=-\bar\omega^\wedge\nu-\bar a^\wedge\phi+(Kx)_a+\mathrm{Exp}(\phi)w_a$. The position row is $\dot\rho=-\bar\omega^\wedge\rho+\nu$.
\end{proof}

The closed loop has the form of~\eqref{eq:closedloop} with $\bar A(\bar\omega(t),\bar a(t))+B_cK$ in place of $A(t)+K$ and $x$ in place of $\xi$. The disturbance of the attitude row splits as in Section 3.1 into the input, the bilinear term $\tfrac12\phi\times w_g$ and a remainder. The disturbance of the velocity row splits into the input, the bilinear term $\phi\times w_a$ and a remainder. Lemma~\ref{lem:remainder} bounds the remainder of the attitude row, and the closed form of the exponential map bounds the remainder of the velocity row, which is of second order in the attitude error, on the a priori enclosure with the attitude part $R_\phi$ of the half-widths $R$ by
\[
  |(\mathrm{Exp}(\phi)-I-\phi^\wedge)w_a|\le\Bigl[\bigl(1-\tfrac{\sin\theta_R}{\theta_R}\bigr)|R_\phi^\wedge|+\tfrac12|R_\phi^\wedge|^2\Bigr]\sigma_a\mathbf 1,\quad \theta_R=\|R_\phi\|.
\]
Algorithm~\ref{alg:tube} and Theorem~\ref{thm:soundness} therefore apply to these vehicles.

The bilinear term $\phi\times w_a$ of the velocity row grows with the attitude error, so a larger a priori enclosure gives a larger bound on this term, which in turn enlarges the enclosure. Under large disturbances, this loop can make the enclosure grow without bound. We therefore enclose the rotated disturbance $\mathrm{Exp}(\phi)w_a$ as a whole. A rotation preserves the norm, so $\mathrm{Exp}(\phi)w_a$ lies in the ball of radius $\sqrt3\,\sigma_a$ for every attitude error, and a zonotope that contains this ball encloses it with inputs that enter linearly.

\begin{lemma}[Ball enclosure]\label{lem:ball}
There is a zonotope $\mathcal Z_b=\langle0,G_b\rangle$ with twelve generators that contains the ball of radius $\sqrt3\,\sigma_a$. For every $\phi\in\mathbb R^3$ and every measurable $w_a(t)$ with $|w_a|\le\sigma_a\mathbf 1$, there is a measurable $\beta(t)\in[-1,1]^{12}$ with $\mathrm{Exp}(\phi(t))w_a(t)=G_b\,\beta(t)$.
\end{lemma}

\begin{proof}

We construct $\mathcal Z_b$ from twelve unit vectors and show that its inradius is at least $\sqrt3\,\sigma_a$. Let $r_b=\sqrt3\,\sigma_a$, let $g^{(1)},\dots,g^{(12)}\in\mathbb R^3$ be pairwise non-parallel unit vectors that do not lie in one plane, let $n_{ij}$ be the normalized cross product of $g^{(i)}$ and $g^{(j)}$, and let
\begin{equation}\label{eq:inradius}
  s=\min_{i<j}\sum_{l=1}^{12}|g^{(l)\top}n_{ij}|,\qquad G_b=c\,[g^{(1)}\ \cdots\ g^{(12)}],\quad c\ge r_b/s .
\end{equation}
The zonotope $\mathcal Z_b$ is centrally symmetric with facet normals $\pm n_{ij}$, so it is the intersection of the slabs $|n_{ij}^\top y|\le c\sum_l|g^{(l)\top}n_{ij}|$, each of half-width at least $cs\ge r_b$. Hence $\mathcal Z_b$ contains the ball of radius $r_b$, which contains $\mathrm{Exp}(\phi)w_a$ because $\|\mathrm{Exp}(\phi)w_a\|=\|w_a\|\le r_b$. The set of $\beta$ with $G_b\beta=\mathrm{Exp}(\phi(t))w_a(t)$ is nonempty and compact and depends measurably on $t$, so a measurable selection exists~\cite{kuratowski1965selectors}.
\end{proof}

The velocity row thus has two inputs. The box input is $w_a$ with $|w_a|\le\sigma_a\mathbf 1$, for which the rotated disturbance splits into $w_a$, the bilinear term $\phi\times w_a$ and a remainder. The ball input is $\beta\in[-1,1]^{12}$ with $\mathrm{Exp}(\phi)w_a=G_b\beta$, for which the velocity row has neither a bilinear term nor a remainder. The price of the ball input is a larger input set, because the box of $w_a$ is inscribed in the ball. Replacing $\mathrm{Exp}(\phi)w_a$ by $G_b\beta$ in Proposition~\ref{prop:hybrid} gives a closed loop of the form~\eqref{eq:closedloop}, in which the velocity row has the input matrix $G_b$ and the disturbance $\beta$ with half-widths $\mathbf 1$, so both inputs give valid tubes. In every step, the tube propagates the current set with both inputs, discards an input whose a priori enclosure fails, and keeps the result with the smaller attitude bound.

\begin{figure*}
\centering
\includegraphics[width=\textwidth]{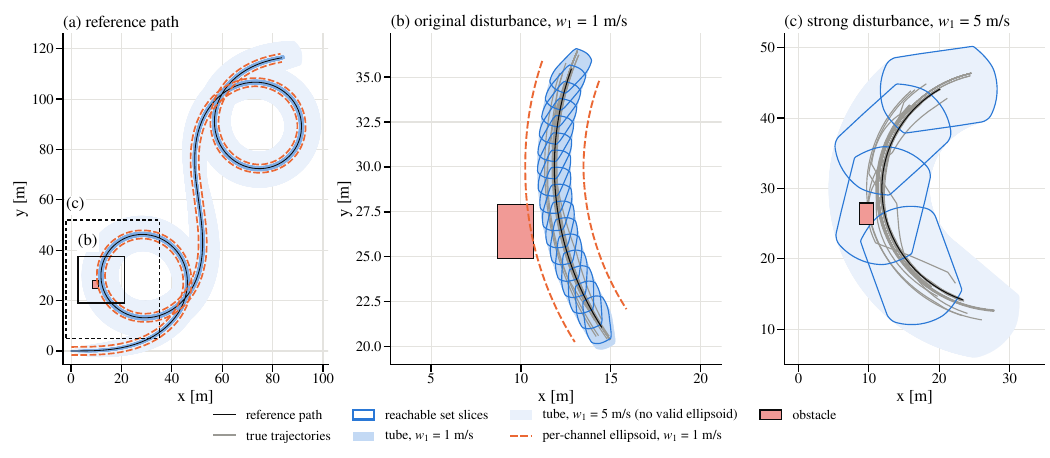}
\caption{Tubes of the planar aircraft in the world frame around an obstacle near the reference path. (a) The whole path. (b) Under the original disturbance, the tube clears the obstacle, whereas the region swept by the ellipsoid overlaps it. (c) Under the strong disturbance, the tube overlaps the obstacle, and so do trajectories of the true closed loop. No ellipsoid is valid.}\label{fig:tube}
\end{figure*}

\section{Experiments}

\subsection{Setup}

We evaluate on three simulated systems and on a real quadrotor. On the planar aircraft, we compare the tube with the ellipsoid method~\cite{lin2024loglinear}, and on the attitude benchmark with the interval method~\cite{harapanahalli2024}. The multirotor tests the hybrid coordinates, and flights of the quadrotor test the tube on a real vehicle.

We bound the attitude error angle and, where a position exists, the distance between the positions of the vehicle and the reference. The tube uses the step $h=0.01$ s on the simulated systems and $h=0.005$ s in the flights. The bounds are read from its time-interval sets, so they hold at every time of the horizon. The true reachable set is not available. On the simulated systems, a lower bound on each quantity therefore comes from Monte Carlo trajectories and from worst-case trajectories of the true closed loop, whose disturbance is chosen along the costate direction. The ratio of the tube bound to the lower bound at the final time $T$ measures the tightness of the tube.

On the planar aircraft, the multirotor and the quadrotor, we compare with two forms of the ellipsoid method. The joint-bound ellipsoid is the invariant ellipsoid of the ellipsoid method~\cite{lin2024loglinear}, which bounds all disturbance components jointly by one norm bound. The per-channel ellipsoid is our variant of the same method with one multiplier per disturbance channel. Both replace the matrices that multiply the disturbance in the closed loop by bounds that hold on the whole ellipsoid. These matrices are the inverse right Jacobian and, on the multirotor, $\mathrm{Exp}(\phi)$. We solve both on a grid of the free parameter of the linear matrix inequality and report the best value for each quantity. All methods use the same feedback gain. The bounds of the tube and of the two ellipsoids are guaranteed bounds. Computation times are wall-clock times of single runs.

\subsection{Planar Aircraft and Attitude Benchmark}

The planar aircraft uses the kinematic model on $SE(2)$ and the initial box of the example of the ellipsoid method. It tracks a reference whose speed and turn rate vary between steps and are constant within each step over the horizon $T=20$ s with a linear quadratic regulator. The disturbances of the example are at most $w_1$ m/s on the two velocity components and $0.1$ rad/s on the turn rate. We use the two levels $w_1=1$ m/s and $w_1=5$ m/s and call them the original and the strong disturbance. On this system, we run Algorithm~\ref{alg:tube} in two variants. The variant with a fixed step uses the step $h$. The variant with refined steps first takes a trial step of length $h$ and, when the discretization error of this step exceeds a tolerance, divides the step into up to 16 sub-steps and applies Algorithm~\ref{alg:tube} to each sub-step.

Under the original disturbance, the tube with refined steps bounds the heading error $\theta$ by 0.131 rad and the position error by 1.28 m. These bounds are 51\,\% and 20\,\% smaller than those of the per-channel ellipsoid, the tighter of the two ellipsoids. At $T$, the bounds of the tube are within 10\,\% of the lower bounds. Under the strong disturbance, the tube bounds the two errors by 0.757 rad and 8.04 m, and its bounds at $T$ are within 42\,\% of the lower bounds. No ellipsoid is valid under this disturbance. The ratio to the lower bound grows with the disturbance because the bilinear term grows with the error. Table~\ref{tab:1} lists the bounds and the computation times of both variants. The variant with a fixed step is 10 to 24 times faster, and its bounds are about 2\,\% larger under the original disturbance and about 12\,\% larger under the strong disturbance.

\begin{table*}
\caption{Maximum over $[0,T]$ of the bounds on the heading error and on the position error of the planar aircraft, with the computation time. The ratios to the lower bound in the text are taken at the final time $T$.}\label{tab:1}
\begin{tabular}{lrrrrrr}
\toprule
method & $w_1=1$ heading [rad] & $w_1=1$ position [m] & time & $w_1=5$ heading [rad] & $w_1=5$ position [m] & time \\
\midrule
lower bound & 0.112 & 1.15 & --- & 0.496 & 5.71 & --- \\
tube, refined steps & 0.131 & 1.28 & 12 s & 0.757 & 8.04 & 29 s \\
tube, fixed step & 0.134 & 1.30 & 1.2 s & 0.849 & 8.96 & 1.2 s \\
per-channel ellipsoid & 0.268 & 1.59 & 4.2 s & \multicolumn{2}{c}{no valid set} & 2.0 s \\
joint-bound ellipsoid & 2.28 & 10.7 & 32 s & \multicolumn{2}{c}{no valid set} & 0.9 s \\
\bottomrule
\end{tabular}
\end{table*}

Figure~\ref{fig:tube} shows the tubes in the world frame with one obstacle at a distance of 1.3 m from the reference path. Under the original disturbance, the tube clears the obstacle with a margin of 0.37 m, whereas the region swept by the per-channel ellipsoid overlaps it. Only the tube therefore certifies that this trajectory avoids the obstacle. The ellipsoid is one set for the whole trajectory, so the region that it sweeps along the reference path keeps nearly the same lateral width. The lateral width of the tube follows the turn rate of the reference, because the heading uncertainty rotates the longitudinal uncertainty into the lateral direction at this rate. Under the strong disturbance, the tube overlaps the obstacle. This overlap reflects a real hazard, since trajectories of the true closed loop enter the obstacle by up to 0.96 m.

The attitude benchmark of the interval method~\cite{harapanahalli2024} evolves on $SO(3)$ with a given body rate $\omega(t)$ and no feedback over $T=5$ s, and we use its disturbance bound and initial box. On this benchmark, the tube bounds the attitude error angle at $T$ by 0.106 rad in 0.28 s, which is 1.07 times the lower bound. The interval method bounds it by 1.61 rad in 0.15 s, which is 16.2 times the lower bound. The undisturbed error dynamics $\dot\xi=-\omega(t)^\wedge\xi$ rotate the set of tracking errors. The box of the interval method therefore grows by the wrapping effect, whereas the zonotope rotates with the dynamics without enlargement.

\subsection{Multirotor}

The angular rate and the specific force of the fully actuated multirotor act in every direction, while its position error is driven only through the velocity. The multirotor starts from a small box of initial tracking errors and tracks a smooth yaw-dominated reference maneuver over $T=20$ s with a linear quadratic regulator. Its disturbances have two levels. The nominal level has $\sigma_g=0.05$ rad/s and $\sigma_a=0.5$ m/s$^2$, and the strong level has three times these values. The ellipsoid is conservative because it bounds all disturbance components jointly by one norm ball and because it keeps one size for the whole trajectory. To separate the two sources, we also construct a time-varying norm bound for comparison. It propagates the same time-varying linear part as the tube but replaces the disturbance terms by a ball whose radius bounds their norm and depends on the attitude error. It thus keeps the time variation of the tube and loses the componentwise bounds of the disturbance.

\begin{figure}[!t]
\centering
\includegraphics[width=\columnwidth]{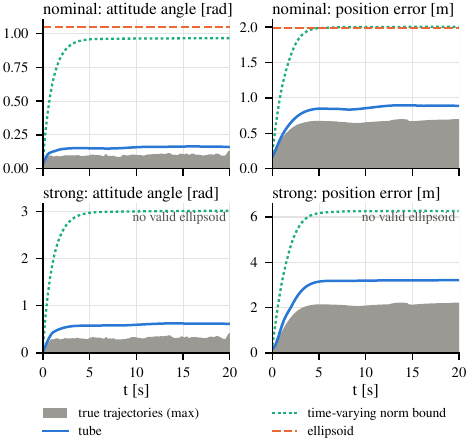}
\caption{Bounds on the attitude error and the position error of the multirotor under nominal and strong disturbances. The tube lies just above the largest error of the true closed loop and far below the ellipsoid and the time-varying norm bound. Under the strong disturbances no ellipsoid is valid.}\label{fig:bounds}
\end{figure}

In hybrid coordinates, the tube bounds the attitude error by 0.167 rad and the position error by 0.898 m under nominal disturbances. These bounds are 84\,\% and 55\,\% smaller than those of the ellipsoid. Under the strong disturbances, the tube remains valid, whereas no ellipsoid is valid. Table~\ref{tab:2} lists the bounds, and Figure~\ref{fig:bounds} shows them over time.

\begin{table*}
\caption{Maximum over $[0,T]$ of the bounds on the attitude error angle and on the position error of the multirotor, with the computation time. The ellipsoid is the joint-bound ellipsoid, which gives the smaller bounds. In every step, the tube keeps the box input or the ball input of Lemma~\ref{lem:ball}, whichever gives the smaller attitude bound.}\label{tab:2}
\setlength{\tabcolsep}{2.5pt}\begin{tabular}{lrrrrrr}
\toprule
method & nominal, attitude [rad] & nominal, position [m] & time & strong, attitude [rad] & strong, position [m] & time \\
\midrule
lower bound & 0.134 & 0.701 & --- & 0.419 & 2.22 & --- \\
tube & 0.167 & 0.898 & 12 s & 0.630 & 3.21 & 34 s \\
ellipsoid & 1.05 & 1.99 & 1.9 min & no valid set &  & 44 s \\
time-varying norm bound & 0.964 & 2.00 & 3.0 min & 3.02 & 6.27 & 3.0 min \\
\bottomrule
\end{tabular}
\end{table*}

Most of the gain over the ellipsoid comes from the componentwise bounds of the disturbance. The time-varying norm bound reduces the attitude bound of the ellipsoid by 8\,\% and gives a position bound within 1\,\% of the ellipsoid bound. Under the strong disturbances, the time-varying norm bound remains valid, but its attitude bound comes close to the limit $\pi$ of the coordinates.

With the box input alone, the a priori enclosure leaves the domain $\theta<\pi$ at 4.9 s under the strong disturbances. The box input gives the smaller step while the attitude error is small, because the ball input encloses the rotated disturbance in a larger set. As the attitude error grows, the bilinear term and the remainder of the box input in the velocity row grow with it, and the ball input gives the smaller step. Under nominal disturbances, the tube keeps the box input in every step. Under strong disturbances, it keeps the box input for the first 0.87 s and the ball input afterwards, and it certifies the whole horizon.

Without the hybrid coordinates, neither the tube nor any ellipsoid certifies this closed loop. In Lie algebra coordinates, the position row keeps the unmatched term. The a priori enclosure then leaves the domain $\theta<\pi$ after 1.2 s under nominal disturbances, and all ellipsoids are infeasible or reach $\theta=\pi$.

\subsection{Ablation}

The tubes on the planar aircraft and the multirotor need both the remainder bound of Lemma~\ref{lem:remainder} and their coordinates, which are Lie algebra coordinates on the planar aircraft and hybrid coordinates on the multirotor.

First, we keep the coordinates and replace Lemma~\ref{lem:remainder} by a generic remainder bound, the Lagrange remainder from an interval evaluation of the Hessian. Under the original disturbance of the planar aircraft and the nominal disturbances of the multirotor, the tube at the final time is then 1\,\% to 7\,\% wider. Under the strong disturbances, it fails at about 7 s on the planar aircraft and at about 2 s on the multirotor. The generic bound grows faster with the box than the bound of Lemma~\ref{lem:remainder}, so the a priori enclosure stops closing.

Second, we also replace the coordinates by the world-frame position, velocity and Euler angles and keep the generic bound. This Cartesian variant is the conservative linearization of standard reachability analysis~\cite{althoff2008}. It fails within the first second on both systems at both disturbance levels, because in Cartesian coordinates the true remainder itself grows with the heading and attitude errors.

\begin{figure}[!t]
\centering
\includegraphics[width=\columnwidth]{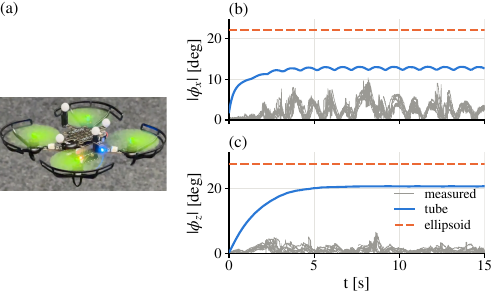}
\caption{(a) The quadrotor. (b), (c) Magnitudes of the roll and yaw components of the attitude error measured in eight flights, with the bounds of the tube and the ellipsoid. The tube contains every measurement and is tighter than the ellipsoid.}\label{fig:hardware}
\end{figure}

\subsection{Flight Experiment on a Quadrotor}

The vehicle is a Crazyflie 2.1 Brushless tracked by a motion capture system. The attitude loop tracks a reference on $SO(3)$ that oscillates in roll and pitch and rotates in yaw. A position controller adds a small tilt that keeps the vehicle near its starting point. The attitude law is the dynamic inversion of Section 3.1 on $SO(3)$. The body rate commanded by this law is sent to the onboard rate loop. The disturbance $w$ is the difference between the body rate measured by the gyroscope and the rate commanded to the onboard loop, including the tilt of the position controller. We obtain the disturbance bound and the initial box from 10 calibration flights of the same maneuver. The bound $\sigma$ is the largest $|w|$ per axis over these flights, and the initial box contains the attitude errors of these flights at the start of the reference.

In the eight evaluation flights, the attitude errors measured in all 14099 frames lie in the tube. None of the disturbance samples exceed $\sigma$. The tube bounds are 25\,\% to 41\,\% smaller than the tighter of the two ellipsoids per axis. Figure~\ref{fig:hardware} shows the measured roll and yaw errors of the eight flights with the tube and the ellipsoid bounds. In roll and yaw, the largest tube bounds are $13.1^\circ$ and $20.8^\circ$, the ellipsoid bounds are $22.1^\circ$ and $27.7^\circ$, and the largest measured errors are $10.4^\circ$ and $6.6^\circ$. The yaw bound is the least tight. About half of its width comes from the linear part and two-fifths from the bilinear term $\tfrac12\phi\times w$. The yaw component of this term is driven by the roll and pitch disturbances, which are the largest of the three.
\FloatBarrier
\section{Conclusion}

We presented a zonotope-based reachability algorithm that certifies trajectory tracking on matrix Lie groups by bounding the tracking error of group-affine systems in Lie algebra coordinates. The algorithm propagates the set of tracking errors with the linear part of the error dynamics, accumulates the disturbance in new generators, and bounds the bilinear term and the remainder on an a priori enclosure of each time step. Its tube contains the tracking error at every time instant. When some error components have no input, the closed loop has unmatched nonlinear terms. Hybrid coordinates make these terms matched, so a static feedback cancels them and the same algorithm certifies the closed loop. In every comparison on a planar aircraft, an attitude benchmark, a fully actuated multirotor and a real quadrotor, the tube bounds are smaller than the bounds of the ellipsoid method and of the other methods. The tubes remain valid under disturbances for which no ellipsoid is valid for the same feedback gain. They contain every simulated and measured tracking error.

\bibliographystyle{ACM-Reference-Format}
\bibliography{references}

\end{document}